\documentclass[aps,prd,epsf,amsmath,amssymb,graphics,nofootinbib,preprint]{revtex4}
\usepackage{graphicx}% Include figure files
\usepackage{dcolumn}% Align table columns on decimal point
\usepackage{bm}% bold math

\usepackage{color}

\newcommand{\ts}{t_{\rm s}}
\newcommand{\rs}{r_{\rm s}}
\newcommand{\etas}{\eta_{\rm s}}
\newcommand{\chis}{\chi_{\rm s}}

\newcommand{\as}{a_{\rm s}}
\newcommand{\fs}{f_{\rm s}}

\newcommand{\etac}{\eta_{\rm c}}

\newcommand{\TGH}{T_{\rm GH}}

\begin{document}

\title{Radiative gravastar with thermal spectrum\\
{\it -- Sudden vacuum condensation without gravitational collapse --}}

\author{Ken-ichi Nakao$^{1,2}$}
\author{Kazumasa Okabayashi$^3$}
\author{Tomohiro Harada$^4$}

\affiliation{
${}^{1}$
Department of Physics, Graduate School of Science, Osaka Metropolitan University, 3-3-138 Sugimoto, Sumiyoshi, Osaka City 558-8585, Japan\\
${}^{2}$Nambu Yoichiro Institute of Theoretical and Experimental Physics,
Osaka Metropolitan University, 3-3-138 Sugimoto, Sumiyoshi, Osaka City 558-8585, Japan\\
${}^{3}$Yukawa Institute for Theoretical Physics, Kyoto University, Kitashirakawa Oiwakecho, Sakyo-ku, Kyoto, 606-8502, Japan\\
${}^{4}$Department of Physics, Rikkyo University, Toshima, Tokyo 171-8501, Japan}

%%%%%%%%%%%%%%%%%%%%%%%%%%%%%%%%%%%%%%%%%
%%%%%%%%%%%%%%%%%%%%%%%%%%%%%%%%%%%%%%%%%

\begin{abstract}
The gravastar is an exotic compact object proposed as a final product of gravitational collapse of a massive object 
in order to resolve problems associated with black holes. It is enclosed by a thin crust and the inside of it is occupied 
by the positive cosmological constant. Recently, the present authors studied quantum particle creation 
through spherically symmetric gravitational collapse to form a gravastar, 
and showed that the newly formed gravastar emits thermal radiation with the Gibbons-Hawking temperature of  its de Sitter core. 
In this paper, in order to understand more about the thermal radiation associated with the 
gravastar formation, we investigate the quantum particle creation in another toy model of the gravastar formation; 
a star with the hollow inside suddenly becomes a gravastar through gravitational vacuum condensation. % without gravitational collapse. 
We find that the thermal radiation is emitted from the gravastar just formed also in the present model. 
The thermal radiation from the gravastar just formed comes from the change of the geometry inside the star 
accompanied by gravitational vacuum condensate.  
\end{abstract}

\preprint{AP-GR-186}
\preprint{NITEP 150}
\preprint{YITP-22-141}
\preprint{RUP-22-24}
\date{\today}
\maketitle

%\newpage
%%%%%%%%%%%%%%%%%%%%%%%%%%%%%
%%%%%%%%%%%%%%%%%%%%%%%%%%%%%
%%%%%%%%%%%%%%%%%%%%%%%%%%%%%
\section{Introduction}
%%%%%%%%%%%%%%%%%%%%%%%%%%%%%
%%%%%%%%%%%%%%%%%%%%%%%%%%%%%
%%%%%%%%%%%%%%%%%%%%%%%%%%%%%

As well known, Hawking made a prediction leading to the present main stream 
of the research for the quantum aspect of gravity through theoretical study of the free quantum field 
in the spacetime  with a gravitational collapse of a massive object which will form a black hole; even if a quantum field 
is initially in a vacuum state, the asymptotic observer will observe the quantum radiation 
with thermal  spectrum in the causal future of the late stage of the gravitational collapse 
\cite{Hawking:1974,Hawking:1975} (see for example Ref.~\cite{Pad:2010} for the research of the quantum aspect of gravity) .
This is the so-called Hawking radiation which  
plays a key role in the black hole thermodynamics: 
the black hole was regarded as a kind of the thermodynamic system since 
it is made up from the matter and the radiation with enormous degrees of freedom but is completely specified 
by a few parameters like thermodynamic quantities, i.e., the entropy, temperature, etc.  
Furthermore, there are relations between the parameters 
to specify the stationary black hole solution which can be regarded as the 0th, 1st and 2nd laws of the thermodynamic system 
by identifying the temperature of the Hawking radiation with that of the black hole (see for example Ref.~\cite{Wald}).  

Paranjape and Padmanabhan showed that the shrinking spherical thin null shell causes the quantum radiation 
whose radiation power asymptotically approaches the value of the Hawking radiation even without the formation of the black hole \cite{PP:2009}. 
Barcel\'{o} et al revealed the minimal condition for the thermal spectrum of the quantum radiation caused by the gravitational 
collapse \cite{BLSV:2011}. This minimal condition implies that the spectrum of the quantum radiation studied 
by Paranjape and Padmanabhan asymptotically approaches the thermal one\cite{PP:2009}. 
Thus we may conclude from their results that the thermal spectrum of the quantum radiation caused by the gravitational collapse 
is not related to the horizon or equivalently the black hole formation even if  
the black hole is the thermodynamic system. The origin of the thermal spectrum comes from the existence of the collapsing domain.  
The outside observers cannot detect any physical influences since the black hole is defined as 
a complement of the causal past of the future null infinity\cite{Penrose:1969,Hawking:1970}. 
We can not observationally know that it is the black hole, as long as it is a black hole \cite{Cardoso-P,Nakao-YH,OAN}. 
The quantum effect is not an exception if it follows the causality. 
  
Recently, the present authors studied the quantum radiation caused by the gravitational collapse of a homogeneous 
dust sphere to form not a black hole but a gravastar \cite{NOH:2022} (hereafter we call this reference Paper I). 
The gravastar was proposed by Mazur and Mottola in order to resolve problems 
associated to the black hole (e.g. the information loss problem) \cite{MM2004}. The model studied in Paper I is a small modification of the preceding studies  
\cite{PP:2009,Harada-CM,Kokubu-H,Okabayashi-HN,Okabayashi-D}. However, the very new thing was revealed in Paper I \cite{NOH:2022}: the quantum radiation with the thermal spectrum of the Gibbons-Hawking temperature \cite{Gibbons-Hawking} associated 
to not the total mass of the system but the de Sitter core of the gravastar will be detected by the asymptotic observers. 

The purpose of the present paper is to understand a mechanism to produce the thermal flux with the 
Gibbons-Hawking temperature in the gravastar formation 
by studying a very simple model without the gravitational collapse, since the thermal spectrum seems to come from 
not the gravitational collapse but the change of the medium in the star from the ordinary matter 
to the vacuum energy, i.e., the positive cosmological constant. 

This paper is organized as follows. In Sec.~II. we give a toy model of the gravastar formation through 
sudden gravitational vacuum condensate inside a spherically symmetric star with hollow inside. 
In Sec.~III, we briefly review the quantum particle creation of 
the massless scalar field in the spherically symmetric spacetime and give a formula to relate  
the radiation power of the quantum particle creation with the redshift of the null geodesics along which 
the mode function propagates. In Sec.~IV, we show 
how to calculate the redshift of the radial null geodesics in the present model. 
In Sec.~V, by using formulae given in Sec.~III and Sec.~IV, we obtain the power of the 
quantum radiation in the toy model of the gravastar formation given in Sec.~II. 
Section VI is devoted to summary. 
Two Appendices are given at the end of the paper; in Appendix A, we show the stress-energy-momentum tensor of the thin shell 
obtained in accordance with the Darmois-Israel junction condition, whereas in Appendix B,  we show the constraint on the 
``size" of the crust enclosing the gravastar from the dominant energy condition. In Appendix C, we derive the redshift suffered by a 
null particle moving from the center of the static gravastar to a static observer at infinity, since it is an important 
parameter in our result. 

We adopt the natural unit $c=\hbar=1$.  The Newton's gravitational constant 
and the Boltzmann constant are denoted by $G$ and $k_{\rm B}$, respectively. The sign convention of the metric 
follows the text book written by Wald\cite{Wald}.

%%%%%%%%%%%%%%%%%%%%%%%%%%%%%
%%%%%%%%%%%%%%%%%%%%%%%%%%%%%
%%%%%%%%%%%%%%%%%%%%%%%%%%%%%
\section{A toy model of a gravastar formation due to gravitational vacuum condensation}
%%%%%%%%%%%%%%%%%%%%%%%%%%%%%
%%%%%%%%%%%%%%%%%%%%%%%%%%%%%
%%%%%%%%%%%%%%%%%%%%%%%%%%%%%

As mentioned, we study a toy model of the gravastar formation through sudden gravitational vacuum condensation in  
the vacuum domain enclosed by a spherically symmetric crust, or in other words, the hollow inside of the star.  
The areal radius of the crust is assumed to be always constant denoted by $\rs$ and the gravitational mass of the system is also unchanged 
even when and after the formation of the gravastar.  Hence, the background system looks static for the observer far from the star.  

The inside of the star is assumed to be described by the Robertson-Walker metric;
\begin{equation}
ds^2=a^2(\eta)\left(-d\eta^2+d\chi^2+\chi^2 d\Omega^2\right), \label{RW-metric}
\end{equation}
where $d\Omega^2$ is the round metric. 

The outside of the star is assumed to be vacuum. By Birkhoff's theorem, the outside domain is described 
by the Schwarzschild metric: 
\begin{align}
ds^2&=-f(r)dt^2+\frac{dr}{f(r)}+r^2 d\Omega^2, \label{Sch-metric}
\end{align}
where
\begin{equation}
f(r)=1-\frac{2GM}{r}. 
\end{equation}
The constant $M$ is the total energy of the system which corresponds to the gravitational mass. We assume $M>0$. 

The following inequality should be satisfied so that the timelike Killing vector field in the Schwarzschild domain 
is tangent to the surface of the star; 
\begin{equation}
f(\rs)>0,
\end{equation}
or equivalently, $\rs >2GM$.  

We assume that the crust is infinitesimally thin. 
The world line of an observer rest on the crust is represented 
in the form, $t=\ts(\tau)$, $r=\rs$ in the Schwarzschild domain, and $\eta=\etas(\tau)$, $\chi=\chis(\tau)$ in the Robertson-Walker domain, 
where $\tau$ is the proper time. By definition of the proper time, we have  
\begin{equation}
\as^2\left[\left(\frac{d\etas}{d\tau}\right)^2-\left(\frac{d\chis}{d\tau}\right)^2\right]=1
=\fs\left(\frac{d\ts}{d\tau}\right)^2-\frac{1}{\fs}\left(\frac{d\rs}{d\tau}\right)^2, \label{normalization} 
\end{equation}
where 
\begin{equation}
\as=a(\etas)~~~{\rm and}~~~\fs=f(\rs).
\end{equation}
Note that 
\begin{equation}
\as\chis=\rs
\end{equation} 
holds.  

From the first equality of Eq.~\eqref{normalization}, we have
\begin{equation}
\frac{d\etas}{d\tau}=\frac{1}{\as\sqrt{1-\chis'^2}}, \label{eta-component}
\end{equation}
where a prime represents a derivative with respect to $\etas$.
From the second equality of Eq.~\eqref{normalization} and $\dfrac{d\rs}{d\tau}=0$, we have
\begin{equation}
\frac{d\ts}{d\tau}=\frac{1}{\sqrt{\fs}}. \label{t-component}
\end{equation}

%%%%%%%%%%%%%%%%
%%%%%%%%%%%%%%%%
\subsection{Static phase before gravastar formation}
%%%%%%%%%%%%%%%%
%%%%%%%%%%%%%%%%

Before the star becomes a gravastar, it is static and composed of a single crust 
enclosing a spherically symmetric vacuum domain. By Birkhoff's theorem, the 
vacuum region inside the star is described by the Minkowski geometry, and hence  
the scale factor $a$ in Eq.~\eqref{RW-metric} is constant there. 
We set $a$ to be unity, so that the radial coordinate 
$\chi=\chis$ at the surface of the star is given as
\begin{equation}
\chis=\rs={\rm constant}. \label{surface-ST}
\end{equation} 

As mentioned, the crust is assumed to be an infinitesimally thin shell which can be treated by Darmoise-Israel formalism \cite{Israel:1966}. 
In accordance with Appendix A, and by using Eqs.~\eqref{eta-component} and \eqref{t-component}, 
the energy per unit area, $\sigma$, and the tangential pressure, $p$, of the crust are given as
\begin{align}
\sigma&=\frac{1}{4\pi G\rs}\left(1-\sqrt{\fs}\right), \\
p&=\frac{1}{16\pi G\rs\sqrt{\fs}}\left(1-\sqrt{\fs}\right)^2.
\end{align}
It is easy to see that both $\sigma$ and $p$ are positive. We have
\begin{equation}
\frac{p}{\sigma}=\frac{1-\sqrt{\fs}}{4\sqrt{\fs}}.
\end{equation}
%We assume that the surface stress-energy-momentum tensor satisfies the weak energy condition $\sigma\geq0$. 
It is not difficult to see that the dominant energy condition ($\sigma\geq |p|$) holds if and only if 
\begin{equation}
\fs \geq \frac{1}{25}~~{\rm or}~{\rm equivalently}~~\rs \geq\frac{25}{12}GM. \label{limit-Vac}
\end{equation}

%%%%%%%%%%%%%%%%
%%%%%%%%%%%%%%%%
\subsection{Gravastar phase}
%%%%%%%%%%%%%%%%
%%%%%%%%%%%%%%%%

As mentioned, the final product is assumed to be a spherically symmetric gravastar 
whose inside is occupied by the positive cosmological constant $\Lambda$, so that 
its spacetime geometry is the de Sitter one. 
The scale factor of the de Sitter spacetime in the expanding flat chart is given as 
\begin{equation}
a(\eta)=-\frac{1}{H\left(\eta-\etac\right)}, \label{sfactor-GS}
\end{equation}
where $\etac$ is a constant which will be determined later, and $H:=\sqrt{\dfrac{\Lambda}{3}}$. 
The domain of $\eta$ is less than $\etac$. 
The radial coordinate of the expanding flat chart at the surface of the 
gravastar, $\chi=\chis(\etas)$, should satisfy $a(\etas)\chis =\rs=$constant, and hence we have
\begin{equation}
\chis=-H\rs(\etas-\etac), \label{surface-GS}
\end{equation}
where $\eta=\etas$ is the conformal time on the surface of the gravastar. 

The areal radius of the gravastar should satisfy 
\begin{equation}
1-H^2\rs^2>0
\end{equation}
so that the timelike Killing vector field in the de Sitter domain is tangent to the surface of the gravastar.  

The surface of the gravastar is also a crust which can be approximated by 
an infinitesimally thin shell. By Eqs.~\eqref{eta-component}, \eqref{sfactor-GS} and \eqref{surface-GS}, we have
\begin{equation}
\frac{d\etas}{d\tau}=-\frac{H\left(\etas-\etac\right)}{\sqrt{1-H^2\rs^2}}.
\end{equation}
Then, we obtain
\begin{align}
\frac{d\chis}{d\tau}&=\frac{d\etas}{d\tau}\chis'=\frac{H^2\rs\left(\etas-\etac\right)}{\sqrt{1-H^2\rs^2}}, \\
\frac{d^2\chis}{d\tau^2}&=\frac{d\etas}{d\tau}\left(\frac{d\chis}{d\tau}\right)'=-\frac{H^3\rs\left(\etas-\etac\right)}{1-H^2\rs^2}.
\end{align}
In accordance with the Appendix A, we have, from these results, the energy per unit area, $\sigma$, and the tangential 
pressure, $p$, of the surface of the gravastar as 
\begin{align}
\sigma&=\frac{1}{4\pi G\rs}\left(\sqrt{1-H^2\rs^2}-\sqrt{\fs}\right), \label{sigma} \\
p&=\frac{1}{8\pi G\rs}
\left(\frac{2H^2\rs^2-1}{\sqrt{1-H^2\rs^2}}+\frac{1+\fs}{2\sqrt{\fs}}\right), \label{P}
\end{align}
As pointed out by Visser and Wiltshire \cite{VW:2004}, the reasonable energy conditions are violated  in the case of the 
gravastar whose areal radius, $\rs$, is very close to its gravitational radius $2GM$ (see also Appendix B). 
However, the possibility of such a gravastar may not be excluded, 
since the gravastar was proposed by assuming some unknown framework beyond Einstein's theory, in which the 
effective stress-energy-momentum tensor may not satisfy the reasonable energy conditions. 

%%%%%%%%%%%%%%%%
%%%%%%%%%%%%%%%%
\subsection{Transition phase}
%%%%%%%%%%%%%%%%
%%%%%%%%%%%%%%%%

As mentioned, we assume that the areal radius, $\rs$, of the star is kept constant, even in the transition phase from the star with the hollow inside to 
the gravastar. We assume that the phase transition occurs in a very short period with respect to $\eta$, 
\begin{equation}
-H^{-1}<\etas<-H^{-1}(1-\varepsilon), 
\end{equation}
where $0<\varepsilon\ll1$ holds.  Then, we put  
\begin{equation}
\chis(\etas)=\rs\left[1+\dfrac{1}{2}\varepsilon^{-3}(H\etas+1)^3\left(H\etas+1-2\varepsilon\right)\right],  \label{surface-TR}
\end{equation}
where  we have set 
\begin{equation}
\etac=\frac{1}{2}H^{-1}\varepsilon \label{etac}
\end{equation}
in Eqs.~\eqref{sfactor-GS} and \eqref{surface-GS}. It is not so difficult to see that $\chis$ is a $C^2$ function of $\etas$. 
The scale factor on the surface of the star is given by $\as=\dfrac{\rs}{\chis}$.

%%%%%%%%%%%%%%%%
%%%%%%%%%%%%%%%%
\subsection{Time in the asymptotic region}
%%%%%%%%%%%%%%%%
%%%%%%%%%%%%%%%%

The time for the static observer in the asymptotic region $r\gg 2GM$ agrees with the time coordinate $t$ in Eq.~\eqref{Sch-metric}. The relation 
between $t$ and the proper time $\tau$ of the surface of the star is given as
\begin{equation}
d\tau=\sqrt{\fs}~dt.
\end{equation}
The relation between $\tau$ and the time coordinate $\etas$ on the surface of the star is
\begin{equation}
d\tau=\as\sqrt{1-{\chis'}^2}d\etas,
\end{equation}
and hence 
\begin{equation}
t=\frac{1}{\sqrt{\fs}}\int \as \sqrt{1-\left(\chis'\right)^2}d\etas, \label{time}
\end{equation}
where a prime represents a derivative with respect to $\etas$.

%%%%%%%%%%%%%%%%%%%%%%%%%%%%%
%%%%%%%%%%%%%%%%%%%%%%%%%%%%%
%%%%%%%%%%%%%%%%%%%%%%%%%%%%%
\section{Quantum particle creation}
%%%%%%%%%%%%%%%%%%%%%%%%%%%%%
%%%%%%%%%%%%%%%%%%%%%%%%%%%%%
%%%%%%%%%%%%%%%%%%%%%%%%%%%%%

As in Paper I, we study the quantum dynamics of the massless scalar field $\phi$ in the spherically symmetric 
asymptotically flat spacetime by the semi-classical treatment in which the effect of the quantum field on the classical 
spacetime geometry is ignored. 
By adopting the double null coordinates, the infinitesimal world interval can be written in the following form;
\begin{equation}
ds^2=-h^2(u,v)dudv+r^2(u,v)d\Omega^2, \label{metric-0}
\end{equation}
where $u$ and $v$ are the retarded and the advanced time coordinates, respectively.
In the case of the Minkowski spacetime, $h=1$ and the areal radius $r$ 
agrees with $(v-u)/2$. Hence $u=v$ holds at $r=0$. 
By contrast, in general dynamical cases, the symmetry center $r=0$ is not $v=u$ but 
\begin{equation}
v=F(u). \label{sym-center}
\end{equation} 

In order to study the quantum effect in this spacetime, we consider the massless free scalar field which is the simplest model but 
sufficient for the present purpose. The Lagrangian density is given as
\begin{equation}
{\cal L}=-\frac{1}{2}\sqrt{-g}g^{\mu\nu}(\partial_\mu\phi)\partial_\nu\phi,
\end{equation}
where $g_{\mu\nu}$, $g^{\mu\nu}$ and $g$ are the metric tensor, its inverse and its determinant, respectively. 
The average of radiation power of the massless scalar field due to the quantum effect caused by the spacetime curvature is estimated 
through the expectation value of the stress-energy-momentum tensor $T_{\mu\nu}$ as
\begin{align}
P(u)=\oint_{r\rightarrow\infty} \langle0|(\hat{T}_{uu}-\hat{T}_{vv})|0\rangle r^2d\Omega = \frac{1}{48\pi}\kappa^2(u), \label{P-def}
\end{align}
where  
\begin{align}
\kappa(u):=-\frac{\dfrac{d^2 F(u)}{du^2}}{\dfrac{dF(u)}{du}}, \label{kappa-def}
\end{align}
and we have ignored the total derivative term in Eq.~\eqref{P-def} since it does not contribute the total emission. 
It is known that if the adiabatic condition $|d\kappa/du|\ll\kappa^2$ holds, the spectrum of the radiation is thermal with the temperature\cite{BLSV:2011}
\begin{equation}
k_{\rm B}T=\frac{\kappa}{2\pi}.
\end{equation}

These results are derived by invoking the S-wave approximation in which only the contributions of the spherically symmetric mode function 
is taken into account \cite{Ford-Parker}. The spherically symmetric mode function ${\cal G}(t,r_*)$ behaves as 
\begin{equation} 
{\cal G}=\frac{1}{\sqrt{4\pi\omega}r}e^{-i\omega v}
\end{equation}
in the past null infinity, $u\rightarrow-\infty$, whereas it behaves as
\begin{equation} 
{\cal G}=\frac{1}{\sqrt{4\pi\omega}r}e^{-i\omega F(u)}
\end{equation}
in the future null infinity, $v\rightarrow +\infty$. The angular frequency of ${\cal G}$ in the past null infinity, $\omega_-$, is equal to $\omega$, 
whereas that in the future null infinity, $\omega_+$, is given by
\begin{equation}
\omega_+=\dfrac{d[\omega F(u)]}{du}.
\end{equation}
Hence the redshift, $z$, suffered by the mode function ${\cal G}$ is given as
\begin{equation}
1+z=\frac{\omega_-}{\omega_+}=\left(\frac{dF(u)}{du}\right)^{-1}.
\end{equation}
Thus, the definition of $\kappa$ given as Eq.~\eqref{kappa-def} is rewritten in the form
\begin{equation}
\kappa=\frac{d\ln(1+z)}{du}.
\end{equation}
By estimating the derivative of redshift $z$ of the spherically symmetric mode function ${\cal G}$ 
with respect to the retarded time $u$, we obtain $\kappa$ and 
the radiation power $P$ through Eq.~\eqref{P-def}. 

%%%%%%%%%%%%%%%%%%%%%%%%%%%%%
%%%%%%%%%%%%%%%%%%%%%%%%%%%%%
%%%%%%%%%%%%%%%%%%%%%%%%%%%%%
\section{How to estimate the redshift of the mode function}
%%%%%%%%%%%%%%%%%%%%%%%%%%%%%
%%%%%%%%%%%%%%%%%%%%%%%%%%%%%
%%%%%%%%%%%%%%%%%%%%%%%%%%%%%

The spherically symmetric mode function propagates along a radial null geodesic whose tangent vector is denoted by $k^\mu$ with 
$k^\theta=k^\varphi=0$. The geometric optics approximation leads to 
\begin{align}
\lim_{v\rightarrow+\infty} k_t&=-\omega_+, \\
\lim_{u\rightarrow-\infty}k_t&=-\omega_-.
\end{align}

In the Schwarzschild region, the geodesic equations imply $k_t$ is a negative constant. 
Through the null condition $k^\mu k_\mu=0$ for the radial null, we have
\begin{equation}
k^r=\pm k_t. 
\end{equation} 
Inside the star, the null geodesic equations imply $k_\eta$ is a negative constant. The null condition for the radial null leads to 
\begin{equation}
k_\chi=\pm k_\eta.
\end{equation}

The following orthonormal basis vectors form the rest frame of the surface of the star; 
their components with respect to the coordinate basis in the Schwarzschild region are given as
\begin{align}
e_{(\tau)}^\mu&=\left(\frac{1}{\sqrt{\fs}}, 0,0,0\right), \\
e_{(n)}^\mu&=\left(0,\sqrt{\fs},0,0\right), \\
e_{(\theta)}^\mu&=\left(0,0,\frac{1}{\rs},0\right), \\
e_{(\varphi)}^\mu&=\left(0,0,0,\frac{1}{\rs\sin\theta}\right),
\end{align}
whereas those with respect to the coordinate basis in the Robertson-Walker region are given as 
\begin{align}
e_{(\tau)}^\mu&=\left(\frac{d\etas}{d\tau}, \frac{d\chis}{d\tau},0,0\right)=\frac{d\etas}{d\tau}\left(1,\chis',0,0\right), \\
e_{(n)}^\mu&=\left(\dfrac{d\chis}{d\tau},\dfrac{d\etas}{d\tau},0,0\right)=\frac{d\etas}{d\tau}\left(\chis',1,0,0\right), \\
e_{(\theta)}^\mu&=\left(0,0,\frac{1}{\rs},0\right), \\
e_{(\varphi)}^\mu&=\left(0,0,0,\frac{1}{\rs\sin\theta}\right),
\end{align}
where $\tau$ is the proper time along the surface of the star, a prime represents a derivative with respect to $\etas$. 

The quantities on the crust when the mode function ${\cal G}$ comes into the star are denoted by the characters with a minus as $A_-$, whereas 
those when ${\cal G}$ comes out from the star by the characters with a plus as $A_+$. 
The $\tau$-component of the tangent vector of the ingoing null geodesic along which the ${\cal G}$ propagates 
is estimated as 
\begin{equation}
k_{(\tau)-}:=e_{(\tau)}^\mu k_\mu=-\frac{\omega_-}{\sqrt{\fs}}
\end{equation}
on the Schwarzschild side, whereas it is represented in the form
%\begin{equation}
%k_{(\tau)-}=N_-\left(1+\chi_{\rm s-}'\right)k_\eta,
%\end{equation}
\begin{equation}
k_{(\tau)-}=\frac{d\eta_{\rm s-}}{d\tau}\left(1+\chi_{\rm s-}'\right)k_\eta,
\end{equation}
in the Robertson-Walker side, where, since $k_{\eta+}=k_{\eta-}$ holds, we denote $k_{\eta\pm}$ by $k_\eta$. Hence, we have
%\begin{equation}
%k_\eta=-\frac{\omega_-}{N_-\left(1+\chi'_{\rm s-}\right)\sqrt{\fs}}.
%\end{equation}
\begin{equation}
k_\eta=-\frac{\omega_-}{\dfrac{d\eta_{\rm s-}}{d\tau}\left(1+\chi'_{\rm s-}\right)\sqrt{\fs}}.
\end{equation}
After the null geodesic goes through the center $\chi=0$, it turns to the outgoing, $k_\chi=-k_\eta>0$. Hence, the $\tau$-component of the null tangent 
at the crust is represented as
%\begin{equation}
%k_{(\tau)+}=N_+\left(1-\chi_{\rm s+}'\right)k_\eta=-\frac{\omega_-N_+\left(1-\chi_{\rm s+}'\right)}{N_-\left(1+\chi'_{\rm s-}\right)\sqrt{\fs}}
%\end{equation}
\begin{equation}
k_{(\tau)+}=\frac{d\eta_{\rm s+}}{d\tau}\left(1-\chi_{\rm s+}'\right)k_\eta=-\frac{\omega_-\dfrac{d\eta_{\rm s+}}{d\tau}\left(1-\chi_{\rm s+}'\right)}
{\dfrac{d\eta_{\rm s-}}{d\tau}\left(1+\chi'_{\rm s-}\right)\sqrt{\fs}}
\end{equation}
on the Robretson-Walker side, whereas
 \begin{equation}
k_{(\tau)+}=-\frac{\omega_+}{\sqrt{\fs}}
\end{equation}
on the Schwarzschild side. Thus, we have
\begin{equation}
1+z:=\frac{\omega_-}{\omega_+}=\frac{a_{\rm s+}}{a_{\rm s-}} \sqrt{\frac{\left(1+\chi_{\rm s+}'\right)\left(1+\chi_{\rm s-}'\right)}{\left(1-\chi_{\rm s+}'\right)\left(1-\chi_{\rm s-}'\right)}}.
\label{redshift}
\end{equation}
Equation \eqref{time} implies that
\begin{equation}
dt_{\rm s}=\frac{\as\sqrt{1-{\chis'}^2}}{\sqrt{\fs}}d\etas
\end{equation}
holds on the crust. Then, we have
\begin{align}
\kappa&=\frac{d\ln(1+z)}{du}=\frac{d\ln(1+z)}{dt_{\rm s}}
=\frac{\sqrt{\fs}}{a_{\rm s+}\sqrt{1-\chi_{\rm s+}'^{2}}}\frac{d\ln(1+z)}{d\eta_{\rm s+}} \nonumber \\
&=\frac{\sqrt{\fs}}{a_{\rm s+}\sqrt{1-\chi_{\rm s+}'^2}}\left(\frac{\partial}{\partial\eta_{\rm s+}}
+\frac{d\eta_{\rm s-}}{d\eta_{\rm s+}}\frac{\partial }{\partial \eta_{\rm s-}}\right)\ln(1+z).
\label{del-z}
\end{align}

%%%%%%%%%%%%%%%%%%%%%%%%%%%%%
%%%%%%%%%%%%%%%%%%%%%%%%%%%%%
%%%%%%%%%%%%%%%%%%%%%%%%%%%%%
\section{Quantum radiation due to gravastar formation}
%%%%%%%%%%%%%%%%%%%%%%%%%%%%%
%%%%%%%%%%%%%%%%%%%%%%%%%%%%%
%%%%%%%%%%%%%%%%%%%%%%%%%%%%%

In order to calculate the power \eqref{P-def} of the quantum radiation generated by the gravastar formation, 
we classify the radial null geodesics into the following five classes (see Fig.~1): 
\begin{itemize} 
\item Class I: null geodesics categorized into this class enter and come out from the star in the static phase before gravastar formation.  
\item Class II: null geodesics categorized into this class enter the star static phase before the gravastar formation and come out from the star 
in the transition phase. 
\item Class III: null geodesics categorized into this class enter the star in the static phase before the gravastar formation and come out from the gravastar. 
\item Class IV: null geodesics categorized into this class enter the star in the transition phase and come out from the gravastar. 
\item  Class V: null geodesics categorized into this class enter and come out from the gravastar. 
\end{itemize} 
The quantity $\kappa$ along the null geodesics in each class is given below.

%%%%%%%%%%%%%%%%%%%%%%%%%%%
\begin{figure}[htbp]
 \begin{center}\
 \includegraphics[width=10cm,clip]{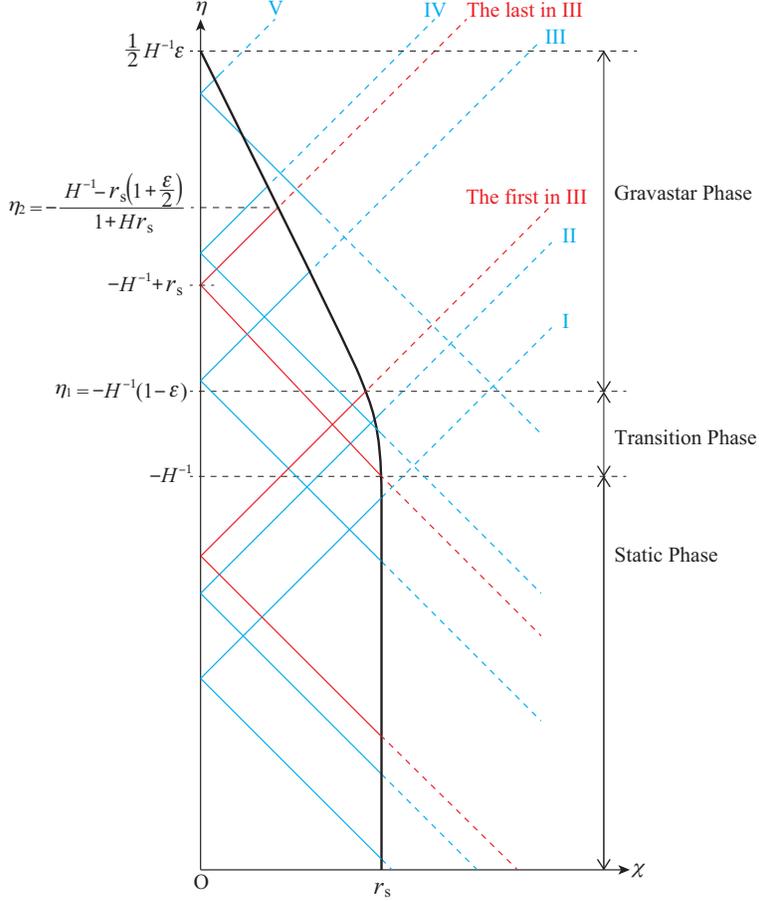}
 \end{center}
 \caption{The schematic spacetime diagram by using the coordinates inside the star is depicted. A thick black curve is the world line of the crust of the star. 
 Blue lines and a red line are null. In the static phase, the star is composed of only an infinitesimally thin crust and the domain enclosed by this crust 
 is vacuum. The red line is the last null geodesic in Class III, which determines the end of the thermal radiation. 
 }
 \label{Situation}
\end{figure}
%%%%%%%%%%%%%%%%%%%%%%%%%%%

%%%%%%
%%%%%%
\subsection{Class I: no radiation phase} 
When the null geodesics of Class I enter and come out from 
the star, both $\chis'$ and $\chis''$ vanishes, and $\as=1$ holds (see Eq.~\eqref{surface-ST}). Hence, $z$ associated with the null of this class identically vanishes, and therefore $\kappa$ also vanishes. No quantum radiation is generated.  

%%%%%%
%%%%%%
\subsection{Class II: 1st burst}
The null geodesics of Class II come out from the star in the period with the very rapid change of geometry due to the gravitational vacuum condensation,  
and hence, as shown below, $\kappa$ can be very large. 
For the transition phase, instead of $\eta_{\rm s+}$, we adopt a normalized time coordinate $\gamma_+$ defined as 
\begin{equation}
 \eta_{\rm s+}=-H^{-1}\left(1-\varepsilon\gamma_+\right), \label{gamma-def}
\end{equation}
where $\gamma_+$ increases from zero to unity. Then, from Eq.~\eqref{surface-TR}, we have
\begin{align}
\chi_{\rm s+}&=\rs\left[1+\dfrac{1}{2}\varepsilon\gamma_+^3(\gamma_+-2)\right],  \label{dchi-0}\\
\chi_{\rm s+}'&=H\rs\gamma_+^2(2\gamma_+-3), \label{dchi-1}\\
\chi_{\rm s+}''&=6\varepsilon^{-1}H^2\rs\gamma_+(\gamma_+-1), \label{dchi-2}
\end{align}
and $a_{\rm s+}=\dfrac{\rs}{\chi_{\rm s+}}$, whereas $\chi_{\rm s-}'=0=\chi_{\rm s-}''$ and $a_{\rm s-}=1$ hold.  
From Eqs.~\eqref{redshift} and \eqref{del-z}, we have
\begin{align}
\kappa&=\frac{\sqrt{\fs}}{a_{\rm s+}\sqrt{1-\chi_{\rm s+}'^2}}\frac{\partial }{\partial\eta_{\rm s+}}\ln\left( a_{\rm s+}\sqrt{\frac{1+\chi_{\rm s+}'}{1-\chi_{\rm s+}'}}\right) 
=\frac{6H^2\rs\gamma_+(\gamma_+-1)\sqrt{\fs}}{\left[1-H^2\rs^2\gamma_+^4(2\gamma_+-3)^2\right]^{3/2}}\varepsilon^{-1} +{\cal O}(\varepsilon^0),
\end{align}
where we have used $a_{\rm s+}=\dfrac{\rs}{\chi_{\rm s+}}$ to obtain a derivative of $a_{\rm s+}$ with respect to $\eta_{\rm s+}$. 
This result implies that the mode function associated with the null geodesics of Class II 
causes very large radiation power $P\propto\varepsilon^{-2}$, i.e., the burst of the quantum radiation.  

%%%%%%
%%%%%%
\subsection{Class III: thermal radiation}

The mode function along the null geodesics in Class III also causes the non-vanishing radiation power. 
In this case, the null geodesic enters the star with hollow inside and comes out from the gravastar. We have 
$a_{\rm s-}=1$, $\chi_{\rm s-}=\rs$, $\chi_{\rm s-}'=0=\chi_{\rm s-}''$, whereas the scale factor $a_{\rm s+}$ is given by Eq.~\eqref{sfactor-GS}, 
and then Eq.~\eqref{surface-GS} leads to $\chi_{\rm s+}'=-H\rs$ and $\chi_{\rm s+}''=0$.  Equation \eqref{redshift} leads to
\begin{equation}
1+z=a_{\rm s+}\sqrt{\frac{1+\chi_{\rm s+}'}{1-\chi_{\rm s+}'}}=-\frac{1}{H(\eta_{\rm s+}-\etac)}\sqrt{\frac{1-H\rs}{1+H\rs}}.
\end{equation}
Substituting this result into Eq.~\eqref{del-z}, we have
\begin{equation}
\kappa=\frac{d\ln(1+z)}{du}=H\sqrt{\frac{\fs}{1-H^2\rs^2}}=\frac{H}{\alpha}, 
\end{equation}
where $\alpha$ is a positive parameter defined as 
\begin{equation}
\alpha=\sqrt{\frac{1-H^2\rs^2}{\fs}}. \label{alpha-def}
\end{equation}
This result implies that $\kappa$ is exactly constant, 
and hence the adiabatic condition $|d\kappa/du|\ll\kappa^2$ is trivially satisfied. 
The spectrum of the radiation is thermal with the temperature
\begin{equation}
T=\TGH \sqrt{\frac{\fs}{1-H^2\rs^2}}=\frac{\TGH}{\alpha}. \label{temperature}
\end{equation}
where 
\begin{equation}
k_{\rm B}\TGH=\frac{H}{2\pi}
\end{equation}
is the Gibbons-Hawking temperature of the de Sitter spacetime with the Hubble constant $H$ \cite{Gibbons-Hawking}. 
Here it is worth noting that $\alpha$ is equal to the redshift factor suffered by a null particle moving 
from the symmetry center to the static observer at infinity in the static gravastar phase (see Appendix C). 
Hence the temperature $T$ of the quantum radiation can be regarded as the redshifted Gibbons-Hawking temperature $\TGH$. 

From Eqs.~\eqref{sigma} and \eqref{temperature}, we find the relation between the temperature, $T$, and the sign of the energy per unit area, $\sigma$, of the crust as follows; 
\begin{equation}
\begin{cases}
T<\TGH & {\rm for}~\sigma>0, \\
T=\TGH & {\rm for}~\sigma=0, \\
T>\TGH & {\rm for}~\sigma<0.
\end{cases}
\end{equation}
It is an intriguing fact that the weak energy condition $\sigma\geq0$ 
imposes an upper limit $\TGH$ on the temperature $T$ of the quantum radiation. 

Let us calculate the duration $\Delta t$ of this thermal emission. The first null of Class III comes out from the star at 
\begin{equation}
\eta=\eta_1:=-H^{-1}(1-\varepsilon). \label{eta_1}
\end{equation}
The conformal time $\eta=\eta_2$ at which the last null of Class III comes out from the star is 
the intersection between the outgoing null 
\begin{equation}
\chi=\eta-(-H^{-1}+\rs)
\end{equation}
and the world line of the surface of the star which is given by Eq.~\eqref{surface-GS} with $\etac=H^{-1}\varepsilon/2$ (see Fig. \ref{Situation}).  
Hence, we have
\begin{equation}
\eta_2=-\frac{H^{-1}-\rs\left(1+\varepsilon/2\right)}{1+H\rs}. \label{eta_2}
\end{equation}
 By using Eq.~\eqref{time}, we have
\begin{align}
\Delta t&=-\sqrt{\frac{1-H^2\rs^2}{\fs}}\int_{\eta_1}^{\eta_2}\frac{d\eta}{H\eta-\varepsilon/2} 
=H^{-1}\sqrt{\frac{1-H^2\rs^2}{\fs}}\ln\left|\frac{(1-\varepsilon/2)(1+H\rs)}{1+\varepsilon/2-H\rs}\right| \nonumber \\
&=H^{-1}\alpha\ln\left|\frac{1+H\rs}{1-H\rs}\right|+{\cal O}(\varepsilon).
\end{align}
Thus, the total energy $E$ of the emitted thermal radiation is given as
\begin{align}
E&=\frac{1}{48\pi}\kappa^2\Delta t=\frac{H}{48\pi}\sqrt{\frac{\fs}{1-H^2\rs^2}}\ln\left|\frac{(1-\varepsilon/2)(1+H\rs)}{1+\varepsilon/2-H\rs}\right| \nonumber \\
&=\dfrac{H}{48\pi\alpha}\ln\left|\dfrac{1+H\rs}{1-H\rs}\right|+{\cal O}(\varepsilon).
\end{align}
If the weak energy condition $\sigma\geq0$ holds, the parameter $\alpha$ is restricted as $\alpha\geq1$, 
and hence $E$ is bounded above by a value of the order $H$. However, if the weak energy condition may not be satisfied, there is no 
upper bound on $E$; in the limit of $\alpha\rightarrow0$, $E$ diverges as the temperature $T$. 

From Eq.~\eqref{alpha-def}, we have $H=\dfrac{1}{\rs}\sqrt{1-\alpha^2 \fs}$ and hence, by using $\rs=\dfrac{2GM}{1-\fs}$, $H$ is rewritten as 
\begin{equation}
H=\frac{(1-\fs)\sqrt{1-\alpha^2\fs}}{2GM}=1.1\times10^{-29}(1-\fs)\sqrt{1-\alpha^2\fs}\left(\frac{M_\odot}{M}\right){\rm J},
\end{equation}
where $M_\odot$ is the solar mass. 

%%%%%%
%%%%%%
\subsection{Class IV: 2nd burst}

The null geodesic of Class IV enters the star in the transition phase and comes out from the gravastar phase. 
 As in the case of the 1st burst, instead of $\eta_{\rm s-}$, we adopt a normalized time coordinate $\gamma_-$ defined as 
\begin{equation}
 \eta_{\rm s-}=-H^{-1}\left(1-\varepsilon\gamma_-\right), \label{gamma-m-def}
\end{equation}
where $\gamma_-$ increases from zero to unity. Then, from Eq.~\eqref{surface-TR}, we have
\begin{align}
\chi_{\rm s-}&=\rs\left[1+\dfrac{1}{2}\varepsilon\gamma_-^3(\gamma_--2)\right],  \label{dchi-m-0}\\
\chi_{\rm s-}'&=H\rs\gamma_-^2(2\gamma_--3), \label{dchi-m-1}\\
\chi_{\rm s-}''&=6\varepsilon^{-1}H^2\rs\gamma_-(\gamma_--1), \label{dchi-m-2}
\end{align}
and $a_{\rm s-}=\dfrac{\rs}{\chi_{\rm s-}}$. The null geodesic of Class IV which enters the star at $\eta=\eta_{\rm s-}$ comes out from the 
star at 
\begin{align}
\eta_{\rm s+}&=\frac{1}{1+H\rs}\left(\eta_{\rm s-}+\chi_{\rm s-}+\frac{1}{2}\rs\varepsilon\right) \nonumber  \\
&=\frac{H^{-1}}{1+H\rs}\left\{-1+H\rs+\varepsilon\left[\gamma_-+\frac{1}{2}H\rs(\gamma_-^4-2\gamma_-^3+1)\right]\right\},
\end{align}
where we have used Eqs.~\eqref{surface-GS} and \eqref{etac} in the first equality, and Eqs.~\eqref{gamma-m-def} and \eqref{dchi-m-0} 
in the second equality, and $a_{\rm s+}=\dfrac{\rs}{\chi_{\rm s+}}$. Then, we have
\begin{equation}
\frac{d\eta_{\rm s+}}{d\eta_{\rm s-}}=H\varepsilon^{-1}\frac{d\eta_{\rm s+}}{d\gamma_-}
=\frac{1}{1+H\rs}\left[1+H\rs\gamma_-^2(2\gamma_--3)\right].
\end{equation}
 Equation \eqref{redshift} leads to
\begin{equation}
1+z=\frac{\chi_{\rm s-}}{\chi_{\rm s+}}\sqrt{\frac{1+\chi_{\rm s-}'}{1-\chi_{\rm s-}'}}\sqrt{\frac{1-H\rs}{1+H\rs}}.
\end{equation}  
Then, Eq.~\eqref{del-z} leads to
\begin{align}
\kappa&=\frac{\chi_{\rm s+}}{\rs}\sqrt{\frac{\fs}{1-H^2\rs^2}}\frac{d}{d\eta_{\rm s+}}\left[\ln\chi_{\rm s-}-\ln\chi_{\rm s+}
+\frac{1}{2}\ln(1+\chi_{\rm s-}')-\frac{1}{2}\ln(1-\chi_{\rm s-}')\right] \nonumber \\
&=\frac{\chi_{\rm s+}}{\rs}\sqrt{\frac{\fs}{1-H^2\rs^2}}
\left[\frac{H\rs}{\chi_{\rm s+}}+\frac{d\eta_{\rm s-}}{d\eta_{\rm s+}}
\left(\frac{\chi_{\rm s-}'}{\chi_{\rm s-}}+\frac{\chi_{\rm s-}''}{1-\chi_{\rm s-}'^2}
\right)\right] \nonumber \\
&=\frac{6\gamma_-(1-\gamma_-)H^2\rs\sqrt{\fs(1-H\rs)}}
{\left[1+H\rs\gamma^2(3-2\gamma_-)\right]\left[1-H\rs\gamma_-^2(3-2\gamma_-)\right]^2\sqrt{1+H\rs}}\varepsilon^{-1}+{\cal O}(\varepsilon^0).
\end{align}
This result implies that the mode function propagating along null geodesics of Class IV also causes the radiation power $P=\kappa^2/48\pi$ 
proportional to $\varepsilon^{-2}$, i.e., the burst of the particle creation. 

%%%%%%
%%%%%%
\subsection{Class V: no radiation phase again}
The null geodesics of Class V enter and come out from the star in the gravastar phase. If the null geodesic enters the star at 
$\eta=\eta_{\rm s-}$ and $\chi_{\rm s-}$, it comes out from the star at 
\begin{equation}
\eta=\eta_{\rm s+}=\chi_{\rm s+}+\eta_{\rm s-}+\chi_{\rm s-}.
\end{equation}
Substituting $\eta_{\rm s\pm}=-(H\rs)^{-1}\chi_{\rm s\pm}+\etac$ derived from Eq.~\eqref{surface-GS} into this equation, we have
\begin{equation}
(1+H\rs)\chi_{\rm s+}=(1-H\rs)\chi_{\rm s-}.
\end{equation}
Hence, we obtain
\begin{equation}
\frac{a_{\rm s+}}{a_{\rm s-}}=\frac{\chi_{\rm s-}}{\chi_{\rm s+}}=\frac{1+H\rs}{1-H\rs}.
\end{equation}
From this result and Eq.~\eqref{redshift}, we have $z=0$ as expected. 
The mode function propagating along the null geodesics of Class V causes no radiation power. 

In Fig.~2, we depict the schematic picture of the radiation power as a function of the time. 

%%%%%%%%%%%%%%%%%%%%%%%%%%%
\begin{figure}[htbp]
 \begin{center}\
 \includegraphics[width=9cm,clip]{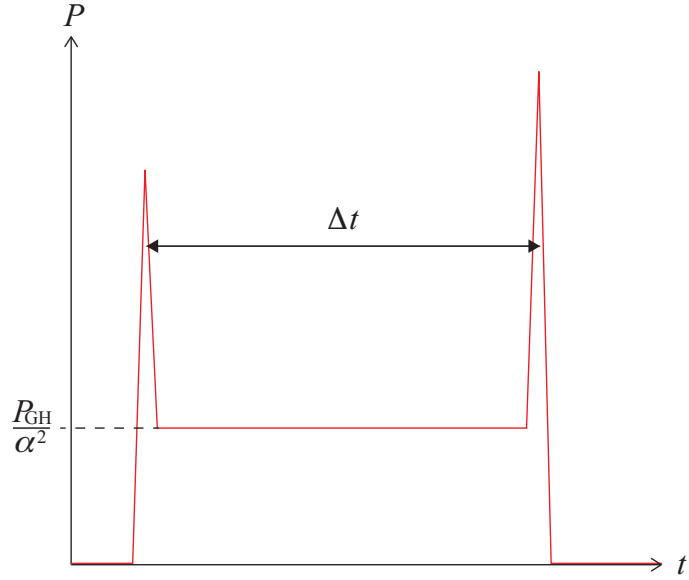}
 \end{center}
 \caption{The schematic diagram for the power $P$ of the quantum radiation as a function of the time for the asymptotic observer. 
Between two bursts of the quantum radiation, there is a duration $\Delta t$ in which the power $P$ is equal to the constant value 
$\alpha^{-2}P_{\rm GH}$, the spectrum of the quantum radiation  is thermal with $T=\alpha^{-1}\TGH$, 
where $P_{\rm GH}=\dfrac{\pi}{12}\left(k_{\rm B} T_{\rm GH}\right)^2$.  
 }
 \label{Power}
\end{figure}
%%%%%%%%%%%%%%%%%%%%%%%%%%%

%%%%%%%%%%%%%%%%%
%%%%%%%%%%%%%%%%%
%%%%%%%%%%%%%%%%%
\section{Summary}
%%%%%%%%%%%%%%%%%
%%%%%%%%%%%%%%%%%
%%%%%%%%%%%%%%%%%

We studied the quantum radiation in the gravastar formation process through the 
vacuum condensation in a spherically symmetric star with constant mass and constant radius. 
The star is assumed to be initially composed of a crust of infinitesimally thin width, which encloses 
a vacuum domain. In other words, the initial situation is a star with hollow inside.  
The inside of the star, i.e., the vacuum domain enclosed by the infinitesimally thin crust
is assumed to suddenly get occupied by the positive cosmological constant, and the star eventually 
becomes a gravastar. By virtue of the spherical symmetry, the initial vacuum domain enclosed by the crust is 
Minkowskian, whereas the inside of the gravastar as a final product has the de Sitter geometry. 

The sudden gravitational vacuum condensation leads two bursts of the particle creation and the thermal radiation 
between these bursts even without the event horizon. 
Therefore the thermal spectrum does not necessarily imply the existence of hidden information. 
In Paper I, we assumed 
\begin{equation}
1-\dfrac{2GM}{\rs}=1-H^2\rs^2 \label{condition}
\end{equation}
in accordance with the original proposal by Mazur and Mottola, and in this case, 
the temperature of the thermal spectrum is equal to the Gibbons-Hawking one.  
On the other hand, in case that Eq.~\eqref{condition} does not hold, the temperature of the thermal radiation is proportional 
to $\sqrt{\dfrac{\fs}{1-H^2\rs^2}}$ which can be regarded as the redshift effect. 
It may be very high in the case of $\fs\gg 1-H^2\rs^2$ 
as well as the total radiated energy $E$, while it is bounded above by $\TGH$ if the weak energy condition 
equivalent to $0< \fs \leq 1-H\rs^2$ is satisfied.  
Although the high temperature needs the violation of the weak energy condition, 
it is not clear whether the weak energy condition holds 
in the process of the gravitational vacuum condensation in 
accordance with the new physics beyond Einstein theory. 

\acknowledgments

KN thanks Satoshi Tanaka for his very significant comment which 
is a motivation of the present study.  KN is also grateful to Hideki Ishihara, Hirotaka Yoshino and 
colleagues at the research group of astrophysics in Osaka Metropolitan University for useful discussions at colloquium. 
This work was supported by JSPS KAKENHI Grant
Number JP21K03557(KN), JP19K03876, JP19H01895, JP20H05853 (TH), and JP21J15676 (KO). 

\appendix

%%%%%%%%%%%%%%%%%
%%%%%%%%%%%%%%%%%
\section{Stress-energy-momentum of infinitesimally thin shell}
%%%%%%%%%%%%%%%%%
%%%%%%%%%%%%%%%%%

The Darmois-Israel junction condition specifies the stress-energy-momentum tensor 
of the infinitesimally thin shell on the surface of the star\cite{Israel:1966}.  

We adopt the orthonormal basis $\bm{e}_{(\alpha)}$ on the surface of the star; we choose $\bm{e}_{(0)}$ is 
the unit tangent of the world line of an observer at rest on the surface of the star. 
Their components with respect to the coordinates just inside the star are given as
\begin{align}
e_{(0)}{}^\mu&=\left(\frac{d\etas}{d\tau},\frac{d\chis}{d\tau},0,0\right) \\
e_{(1)}{}^\mu&=\left(\frac{d\chis}{d\tau}, \frac{d\etas}{d\tau},0,0\right), \\
e_{(2)}{}^\mu&=\left(0,0,\frac{1}{\as\chis},0\right), \\
e_{(3)}{}^\mu&=\left(0,0,0,\frac{1}{\as\chis\sin\theta}\right),
\end{align}
whereas, with respect to the coordinates just outside the star, as
\begin{align}
e_{(0)}{}^{\mu'}&=\left(\frac{d\ts}{d\tau},\frac{d\rs}{d\tau},0,0\right), \\
e_{(1)}{}^{\mu'}&=\left(\frac{1}{\fs}\frac{d\rs}{d\tau},\fs\frac{d\ts}{d\tau},0,0\right), \\
e_{(2)}{}^{\mu'}&=\left(0,0,\frac{1}{\rs},0\right), \\
e_{(3)}{}^{\mu'}&=\left(0,0,0,\frac{1}{\rs\sin\theta}\right).
\end{align}

We define 
\begin{equation}
e^{(\alpha)\mu}=\eta^{(\alpha)(\beta)}e_{(\beta)}{}^\mu~~~~~{\rm and}~~~~~\eta_{(\alpha)(\beta)}e^{(\beta)\mu}=e_{(\alpha)}{}^\mu
\end{equation}
where $\eta^{(\alpha)(\beta)}={\rm diag}[-1,1,1,1]=\eta_{(\alpha)(\beta)}$. 

The components of the second fundamental form $K^{\rm (in)}_{\mu\nu}$ of the surface of the star with respect 
to the coordinates inside the star are given as
\begin{equation}
K^{\rm (in)}_{\mu\nu}=\left(e^{(0)}{}_\mu e_{(0)}{}^\alpha+e^{(2)}{}_\mu e_{(2)}{}^\alpha+e^{(3)}{}_\mu e_{(3)}{}^\alpha\right)\nabla_\alpha e_{(1)\nu}.
\end{equation}
By using this expression, we have the tetrad components of the second fundamental form as 
\begin{align}
K^{\rm (in)}_{(0)(0)}&=-\frac{d\tau}{d\etas}\left(\frac{d^2\chis}{d\tau^2}+\frac{2}{\as}\frac{d\as}{d\etas}\frac{d\etas}{d\tau}\frac{d\chis}{d\tau}\right), \\
K^{\rm (in)}_{(2)(2)}&=\frac{1}{\as}\frac{d\as}{d\etas}\frac{d\chis}{d\tau}+\frac{1}{\chis}\frac{d\etas}{d\tau}=K^{\rm (in)}_{(3)(3)}
\end{align}
and the other components vanish, where we have used Eq.~\eqref{normalization} to rewrite $\dfrac{d^2\etas}{d\tau^2}$ by the derivatives of $\chis$ with respect to $\tau$.
By contrast, the components of the second fundamental form $K^{\rm (out)}_{\mu\nu}$ of the surface of the star with respect to 
the coordinates outside the star are given as
\begin{equation}
K^{\rm (out)}_{\mu'\nu'}=\left(e^{(0)}{}_{\mu'} e_{(0)}{}^{\alpha'}+e^{(2)}{}_{\mu'} e_{(2)}{}^{\alpha'}+e^{(3)}{}_{\mu'} e_{(3)}{}^{\alpha'}\right)\nabla_{\alpha'} e_{(1)\nu'}.
\end{equation}
Then, we obtain the tetrad components of the second fundamental form as
\begin{align}
K^{\rm (out)}_{(0)(0)}&=-\frac{1}{\fs}\frac{d\tau}{d\ts}\left(\frac{d^2\rs}{d\tau^2}+\frac{GM}{\rs^2}\right), \\
K^{\rm (out)}_{(2)(2)}&=\frac{\fs}{\rs}\frac{d\ts}{d\tau}=\frac{1}{\rs}\sqrt{\left(\frac{d\rs}{d\tau}\right)^2+\fs}=K^{\rm (out)}_{(3)(3)}
\end{align}
and all the other components vanish, where we have used Eq.~\eqref{normalization}.
%By using Eqs.~\eqref{etas-dot}--\eqref{rs-ddot}, we concretely obtain $K_{(\alpha)(\beta)}^{\rm (in)}$ and $K_{(\alpha)(\beta)}^{\rm (out)}$. 

The Darmoise-Israel junction condition is
\begin{equation}
K^{\rm (out)}_{(\alpha)(\beta)}-K^{\rm (in)}_{(\alpha)(\beta)}=-8\pi G \left(S_{(\alpha)(\beta)}
-\frac{1}{2}h_{(\alpha)(\beta)}{\rm tr}S\right),
\end{equation}
where $h_{(\alpha)(\beta)}={\rm diag}[-1,0,1,1]$.  Then, the tetrad components, $S_{(\alpha)(\beta)}$, are given as
\begin{align}
S_{(0)(0)}&=-\frac{1}{4\pi G}\left(K^{\rm (out)}_{(2)(2)}-K^{\rm (in)}_{(2)(2)}\right), \\
S_{(2)(2)}&=-\frac{1}{8\pi G}\left(K^{\rm (out)}_{(0)(0)}-K^{\rm (out)}_{(2)(2)}-K^{\rm (in)}_{(0)(0)}+K^{\rm (in)}_{(2)(2)}\right)=S_{(3)(3)}
\end{align}
and the other components vanish. $S_{(0)(0)}$ is the energy per unit area, $\sigma$, whereas $S_{(2)(2)}=S_{(3)(3)}$ 
is the tangential pressure, $p$.

%%%%%%%%%%%%%%%%%
%%%%%%%%%%%%%%%%%
\section{On reasonable energy conditions at the crust of the gravastar}
%%%%%%%%%%%%%%%%%
%%%%%%%%%%%%%%%%%

We show the constraint on the radius coming from the reasonable energy conditions. 
By using the parameter $\alpha$ defined as Eq.~\eqref{alpha-def}, Eqs.~\eqref{sigma} and \eqref{P} are rewritten in the form, 
 \begin{align}
\sigma&=\frac{\sqrt{\fs}}{4\pi G\rs}\left(\alpha-1\right), \label{sigma-1} \\
p&=\frac{2+\alpha+\alpha(1-4\alpha)\fs}{16\pi G\rs\alpha\sqrt{\fs}},
\end{align}
and we obtain
\begin{equation}
\frac{p}{\sigma}=\frac{1}{4\alpha(\alpha-1)}\left[\frac{2+\alpha}{\fs}+\alpha(1-4\alpha)\right].
\end{equation}
The weak energy condition requires $\alpha\geq1$. In this case, we find that the dominant energy condition, $|p|/\sigma \leq1$, holds if and only if
\begin{equation}
\frac{\alpha+2}{\alpha(8\alpha-5)}\leq \fs\leq \frac{\alpha+2}{3\alpha}. \label{D-condition}
\end{equation}
Since the inequality, $0<\fs<1,$ has to hold by its definition, 
we can see from this equation that the dominant energy condition does not hold for $\alpha=1$. 
Hence, the dominant energy condition requires $\alpha>1$, and hereafter we assume this inequality.

From Eqs.~\eqref{alpha-def} and \eqref{D-condition}, we have
\begin{equation}
\frac{\alpha(\alpha+2)}{8\alpha-5}\leq 1-H^2\rs^2\leq \frac{1}{3}\alpha(\alpha+2). \label{D-condition-dS}
\end{equation}
Since $0<1-H^2\rs^2<1$ should hold, the following inequality should be satisfied; 
\begin{equation}
\frac{\alpha(\alpha+2)}{8\alpha-5}<1,
\end{equation}
and this inequality leads to $1<\alpha<5$. The minimum value of the left-most-side of Eq.~\eqref{D-condition-dS} is obtained for
\begin{equation}
\alpha=\alpha_{\rm min}:=\frac{1}{8}\left(5+\sqrt{105}\right).
\end{equation}
Hence we have
\begin{equation}
1-H^2\rs^2\geq \left.\frac{\alpha(\alpha+2)}{8\alpha-5}\right|_{\alpha=\alpha_{\rm min}}=\frac{13+\sqrt{105}}{32}\simeq0.726. \label{bound}
\end{equation}
The constraint $\alpha<5$ and Eq.~\eqref{D-condition} implies
\begin{equation}
\fs>\left.\frac{\alpha+2}{\alpha(8\alpha-5)}\right|_{\alpha=5}=\frac{1}{25}. \label{limit-GS}
\end{equation}
This constraint is the same as that of the crust star case, Eq.~\eqref{limit-Vac}. 
Hence, if the dominant energy condition is strictly satisfied, the areal radius, $\rs$, of the gravastar cannot 
be so small that it plays the role of a black hole mimicker.

%%%%%%%%%%%%%%%%%
%%%%%%%%%%%%%%%%%
\section{The gravitational redshift in the static gravastar phase}
%%%%%%%%%%%%%%%%%
%%%%%%%%%%%%%%%%%

Here we see that the redshift suffered by a null particle from the symmetry center to 
the static observer at infinity in the static gravastar phase is equal to the parameter $\alpha$. 
The tangent of the null geodesic is denoted by $k^\mu$ as in Sec.~IV. 
For simplicity of the calculations, instead of the Robertson-Walker coordinate \eqref{RW-metric}, 
we adopt the static coordinate system for the inside of the gravastar as
\begin{equation}
ds^2=-\left(1-H^2r^2\right)dt_{\rm in}^2+\frac{dr^2}{1-H^2r^2}+r^2d\Omega^2. 
\end{equation}
The areal radius $r$ is common to both the inside of the gravastar and the outside the Schwarzschild domain. 
In this coordinate system, $k_{t_{\rm in}}=-\omega_{\rm e}$ is constant. 
As mentioned in Sec.~IV, in the Schwarzschild domain outside the gravastar, $k_t$ is also constant 
and here denoted by $-\omega_{\rm d}$.

Since the components of the four-velocity of the emitter at the symmetry center $r=0$ is $(1,0,0,0)$, the angular frequency of the 
light at the emitter is equal to $\omega_{\rm e}$. 
The continuity of the null geodesic tangent $k^\mu$ at the crust of the gravastar located at $r=\rs$ leads to 
\begin{equation}
\frac{\omega_{\rm e}}{\sqrt{1-H^2\rs^2}}=\frac{\omega_{\rm d}}{\sqrt{\fs}}.
\end{equation}
Hence, we have the redshift factor as
\begin{equation}
\frac{\omega_{\rm e}}{\omega_{\rm d}}=\sqrt{\frac{1-H^2\rs^2}{\fs}}=\alpha.
\end{equation}
Note that $\sigma=0$ ($\alpha=1$) corresponds to the case with no redshift, $\sigma>0$ ($\alpha>1$) causes the redshift, whereas 
$\sigma<0$ ($\alpha<1$) leads to the blue shift.

\end{document}